\documentclass[onecolumn,preprint,preprintnumbers,amsmath,amssymb]{revtex4}
\usepackage{graphicx}
\usepackage{dcolumn}
\usepackage{multirow}
\usepackage{bm}
\usepackage[colorlinks, dvipdfm, citecolor=blue]{hyperref}
\usepackage{booktabs}

\begin{document}

\title{Replying ``the comment on Interlayer interactions in graphites"}

\author{Xiaobin Chen$^1$, Fuyang Tian$^{2}$, Clas Persson$^{3,4}$, Wenhui Duan$^1$, and Nanxian Chen$^{1,2}$\footnote{Address for communication: nanxian@tsinghua.edu.cn}}

\affiliation{$^1$Department of Physics, Tsinghua University, Beijing 100084,
People¡¯s Republic of China\\
$^2$Institute for Applied Physics, University of
Science and Technology Beijing, Beijing 100083, People¡¯s Republic
of China\\
$^3$Department of Physics, University of Oslo, 0316 OSLO Norway\\
$^4$Department of Materials Science and Engineering, Royal Institute of Technology, 100 44 Stockholm, Sweden
}

\date{\today}

\begin{abstract}
For the moment, there is no exact description of van der Waals (vdW) interactions. ACFD-RPA \cite{Gould1} is expected to better describe vdW bonding, but it is not exact. The PBE/DFT-D2 method is less satisfactory, however, its results are in good agreement with experimental data. Although our fitting technique may weaken (not neglect) the vdW interactions and produce interlayer potentials with weakened vdW, the obtained interlayer potentials reproduce energetics of graphite near the equilibrium interlayer distance very well, as shown in Ref. \cite{Chen}. If having inputs which fully include vdW interactions and having better fitting functions, we believe that interlayer potentials can also fully include vdW interactions in graphite system.
\end{abstract}

\maketitle

\newpage
It is necessary to emphasize that our interlayer potentials are built basing on binding energy curves of AB- and ABC-stacked graphites.
We calculated equilibrium properties including interlayer distance $d_0$, interlayer binding energy $E_b$, and the elastic constants $C_{33}$ for AB-, ABC- and AA-stacked graphites\cite{Chen} using density functional theory. However we chose the $ab~intio$ binding energy curves of AB- and ABC-stacked graphites as inputs. Our choice is based on two reasons: \\
1. Only AB- and ABC-stacked graphites exist in nature; \\
2. Binding energy of ABC-stacked graphite from PBE/DFT-D2 method is slightly larger than that of AB-stacked graphite. This is qualitatively consistent with natural abundance of ABC- and AB-stacked graphite.

In Ref. \cite{Gould}, the authors write

\emph{The poor energetics can be seen most prominently in the case of AA graphite, where
insertion of the AA parameters from Table 2 into Equation 1 of CTPDC[1] gives a potential
well for $E_{graph-AA} = E^{AA}_\phi$ (from Equation 2 of their work) with a depth of 13800meV/Atom
located at $d_0$ = 0.076 \AA...$\phi^{AA}$, leading to a well
of depth 328meV/Atom located at $d_0$ = 1.29~\AA. This is clearly an
unphysical result, and makes portability of the model to new geometries highly dubious.}

 Actually as mentioned in the last sentence in the section of Results in Ref. \cite{Chen}, we have to note the domain of definition of interlayer potential $\phi^\mathrm{AA}(d)$: $d\ge$ 5.2 {\AA}. In our calculation, $d_1=2.6$ {\AA} is the smallest interlayer distance and our algorithm requires that the $d\ge 2d_1= 5.2~$\AA~for $\phi^\mathrm{AA}(d)$. Therefore, simple extrapolation of $\phi^\mathrm{AA}(d)$ to the range of $d<2d_1$ is indeed a misunderstanding to our work, and $\phi^\mathrm{AA}(d)$ at $d=0\sim5.2$ {\AA} (Fig. 1 in Ref. \cite{Gould}) is unnecessary and unreasonable.


Also as shown in Table I in Ref. \cite{Chen}, seven different exchange-correlation functionals are employed in our calculations. For large interlayer distance, $ab~initio$ calculation is not suitable, because vdW interactions might be too weak to be determined. At present, there is no exact value of interlayer distance where vdW interactions can be safely neglected. In order to get best fittings of binding energy curves, vdW interactions are neglected when numeric variation against $d$ is smaller than $10^{-6}$ eV. It is expected that fitting from RSL2 function can ensure good performance near the equilibrium state and also reasonable overall performance. Unfortunately, fitting results still weaken vdW interactions among the range of $6-8$ {\AA}, as shown in Fig. \ref{dos}.
\begin{figure}{\label{dos}}
\includegraphics[scale=0.7]{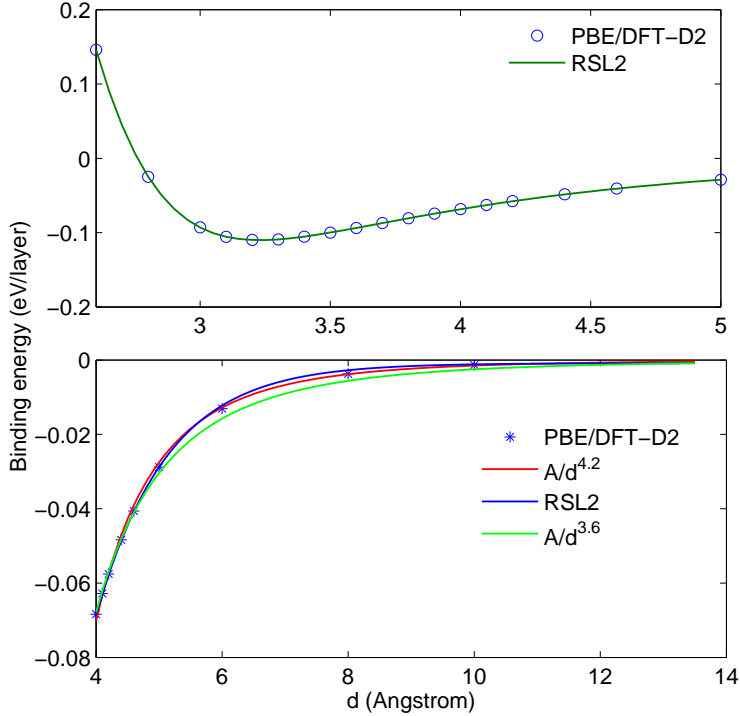}
\caption{The binding energy curve of AB graphite as functions of interlyaer distance. \label{dos}}
\end{figure}

However, it must be noted that the choice of fitting functions either for $E_\mathrm{AB}^{ab \ initio}(d)$ and $E_\mathrm{ABC}^{ab \ initio}(d)$ or for $\phi^\mathrm{AB}(d)$ and $\phi^\mathrm{AA}(d)$ are dependent on both interval domain and sampling distributions. In general, fitting is a typical ill-posed problem. Besides there are some divergence problems of the summations in M\"{o}bius inversion formula especially when the vdW interactions appear.

In Ref. \cite{Gould}, Gould $et$ $al.$ calculate the geometrically determined differences in the potentials energy minima of AB graphite. According to interlayer potentials, we compare the Eq. (2) with Eq. (3) in Ref. \cite{Chen} and get the same $E_\mathrm{Graphite}$ with $E_\mathrm{Exfoliation}$. In Ref. \cite{Bjorkman}, the authors have also mentioned that $E_\mathrm{Exfoliation}$ $\approx$ $E_\mathrm{Graphite}$. We calculated the $E_\mathrm{Bigraphene}-E_\mathrm{Graphite}=1.6$ meV/Atom. The $\Delta E_\mathrm{Bi-Ex}$ from the interlayer potentials is smaller than that of ACDF-RPA in Ref. \cite{Gould}. This difference can be attributed to our $ab$ $initio$ inputs and the weakened vdW interactions caused by fitting. As shown in Fig. \ref{dos}, the binding energy from PBE/DFT-D2 is well fitted by A*$D^{-4.2}$, whereas is poor by the fitting of A*$D^{-3.6}$ which is used in the energy curve reported by Gould $et$ $al.$ \cite{Gould}.

In summary, having the same bigraphene and exfoliation energy do not mean that vdW interactions are neglected in our calculations. It is worth noting that there are lower limits for our interlayer potentials, which are 2.6~\AA~and 5.2~\AA~for $\phi^{\mathrm {AB}}$ and $\phi^{\mathrm {AA}}$, respectively. Also, using less accurate $ab~initio$ results and using one single function to fit interlayer potentials weaken the vdW interacions at large interlayer distance. As pointed out by Gould $et~al.$, our fitting function leads to a too fast decay of binding energy curve $E^{\mathrm{AB}}(d)$.  Nevertheless, we expect that vdW interactions can be correctly included in our interlayer potentials if having better $ab$ $initio$ calculations and better fitting strategy.

Finally, we thank Dr. Gould, Bu\v{c}ko, Leb\`egue, and the Editor for concerns and help to our work.

\begin{small}

\end{small}

\end{document}